\documentclass[prd,aps,showpacs,preprintnumbers,
nofootinbib,notitlepage]{revtex4-1}

\usepackage{amsmath}
\usepackage{amssymb}
\usepackage{graphicx}
\usepackage{epsfig}

\usepackage{epstopdf}
\usepackage{esint}

\begin{document}

\title{Muon decay in orbit: spectrum of high-energy electrons}
\author{Andrzej Czarnecki and Xavier Garcia i Tormo}
\affiliation{Department of Physics, University of Alberta, Edmonton, Alberta,
Canada T6G 2G7}
\author{William J. Marciano}
\affiliation{Department of Physics, Brookhaven National Laboratory, Upton, NY
11973, USA}

\date{\today}

\preprint{Alberta Thy 08-11}

\begin{abstract}
Experimental searches for lepton-flavor-violating coherent muon-to-electron conversion in the field of a nucleus, have been proposed
to reach the unprecedented sensitivity of $10^{-16}-10^{-18}$ per
stopped muon. At that level, they probe new interactions at effective-mass scales well beyond 1000 TeV. However, they must contend with
background from ordinary bound muon decay. To
better understand the background-spectrum shape and rate, we have
carried out a detailed analysis of Coulombic-bound-state muon decay,
including nuclear recoil. Implications for future experiments are
briefly discussed.
\end{abstract}

\pacs{13.35.Bv, 36.10.Ee}

\maketitle

\section{Introduction and Motivation}
From the observation of neutrino oscillations, we now know that lepton
flavors (electron, muon and tau number) are not conserved. However,
the mixing and small neutrino mass differences seen in oscillations
have negligible effect on charged-lepton flavor violating (CLFV) reactions
such as $\mu\to e\gamma$ or $\tau\to\mu\gamma$ (now predicted to occur
with unobservable tiny branching ratios of about $10^{-54}$). So,
if charged-lepton flavor violation were to be experimentally detected,
it would have to come from ``new physics'' such as supersymmetry,
heavy neutrino mixing, leptoquark interactions or some other extension
of the Standard Model. In that way, charged-lepton number violating
reactions provide a discovery window to interactions, beyond Standard
Model expectations, reaching effective-mass scales above
$\mathcal{O}(1000\,{\textrm{TeV}})$ \cite{Kuno:1999jp,LeptMomBook}.

Because muons can be copiously produced at accelerators and are relatively
long lived ($2.2\text{ \ensuremath{\mu}}\mbox{s}$), they have been
at the forefront of searches for CLFV
\cite{Kuno:1999jp,LeptMomBook}.
One reaction that can be probed with particularly high sensitivity
is the muon-electron conversion in a muonic atom, 
\begin{equation}
\mu^{-}+(A,Z)\to e^{-}+(A,Z),\label{eq:mueconv}
\end{equation}
where $(A,Z)$ represents a nucleus of atomic number $Z$ and mass
number $A$. Various experiments have been performed over the years
to search for this process \cite{Marciano:2008zz}. The most recent, and
stringent, results come from the SINDRUM II Collaboration \cite{Bertl:2006up},
which reports an upper limit of $7\times10^{-13}$ for the branching
ratio of the conversion process relative to muon capture in gold, and
a similar unpublished bound for titanium. Several
new efforts are being planned. In the nearest future, the DeeMe
Collaboration \cite{DeeMe} has proposed to reach $10^{-14}$ sensitivity. Larger scale searches,
Mu2e at Fermilab \cite{Carey:2008zz} and COMET at J-PARC \cite{Cui:2009zz},
aim for sensitivities below $10^{-16}$. In the long run, intensity
upgrades at Fermilab and the proposal PRISM/PRIME at
J-PARC may allow them to reach $10^{-18}$ sensitivity, a limit only accessible with
muon-electron conversion in nuclei. For comparison, the
current best upper bound
on the branching ratio of the decay $\mu\to e\gamma$, set by the
MEGA experiment, is $1.2\times10^{-11}$ ($90\%$ confidence level)
\cite{Brooks:1999pu}. For some mechanisms of CLFV, the conversion
process is less sensitive than $\mu\to e\gamma$ by a factor on the order of
a few hundred \cite{Czarnecki:1998iz}. But even in those cases, a $10^{-14}$ conversion search is more sensitive than this best current
bound, and may be competitive with the new search for $\mu\to e\gamma$
by the MEG experiment \cite{Adam:2009ci}. In addition, the conversion
process is also sensitive to CLFV chiral-conserving amplitudes that do not contribute to $\mu\to e\gamma$.

The success of the conversion searches depends critically on control
of the background events. The signal for the $\mu-e$ conversion process
in Eq.~(\ref{eq:mueconv}) is a mono-energetic electron with energy
$E_{\mu e}$, given by 
\begin{equation}
E_{\mu e}=m_{\mu}-E_{\mathrm{b}}-E_{\mathrm{rec}},\label{eq:conven}
\end{equation}
where $m_{\mu}$ is the muon mass, $E_{\mathrm{b}}\simeq Z^2\alpha^2m_{\mu}/2$ is the binding
energy of the muonic atom, and $E_{\mathrm{rec}}\simeq m_{\mu}^2/(2m_N)$ is the nuclear-recoil
energy, with $\alpha$ the fine-structure constant and $m_N$ the
nucleus mass. The main physics background for this signal comes from the so-called
muon decay in orbit (DIO), a process in which the muon decays in
the normal way, i.e. $\mu^{-}\to e^{-}\bar{\nu}_{e}\nu_{\mu}$, while
in the orbit of the atom. Whereas in a free muon decay, in order to
conserve energy and three-momentum, the maximum electron energy is
$m_{\mu}/2$, for decay in orbit the presence of an additional
particle (the nucleus), which can absorb three-momentum, causes the
maximum electron energy to be $E_{\mu e}$. Therefore, the high-energy tail
of the electron spectrum in muon decay in orbit constitutes a
background for conversion searches. A detailed study of that
background is the main focus of this work.

Several theoretical studies of the muon decay in orbit have been performed, starting with Ref.~\cite{Porter:1951zz} about 60
years ago. Reference \cite{Haenggi:1974hp} presented expressions
which allow for a calculation of the electron spectrum including relativistic
effects in the muon wavefunction, the Coulomb interaction between
the electron and the nucleus and a finite nuclear size. Nuclear-recoil
effects which, as we will discuss later, need to be considered in
the high-energy region were only included in the Born approximation
(that is, using the non-relativistic Schr\"odinger wave function for
the muon and a plane wave for the electron), which is not adequate
for the high-energy tail. Later, Refs.~\cite{Watanabe:1987su,Watanabe:1993}
presented similar expressions for the electron spectrum completely neglecting
nuclear-recoil effects, evaluating it for several different elements.
None of these references focused on the high-energy endpoint of the
spectrum, which is the region of interest for conversion experiments.
References \cite{Shanker:1981mi,Shanker:1996rz} did study the high-energy
end of the electron spectrum, and presented approximate results which
allow for a quick rough estimate of the muon decay in orbit contribution
to the background in conversion experiments. However, a detailed evaluation
of the high-energy region of the electron spectrum is still missing
in the literature. What is typically done, to account for
the background from muon decay in orbit, is to connect (in a somewhat arbitrary way) the approximate expressions given in Ref.~\cite{Shanker:1981mi}
with the numerical results presented in Ref.~\cite{Watanabe:1993}.
Since this is the main source of background \cite{Carey:2008zz,Cui:2009zz}
for the oncoming conversion experiments, a more detailed analysis
is highly desirable. In this work we discuss all the relevant effects that need to be included
in the high-energy region of the spectrum and present a precise evaluation
of it. Our results for an aluminum ($Z=13$) nucleus (the intended target in
Mu2e and COMET) are presented in Figure~\ref{fig:speceprec}.

The structure of the paper is as follows. In Sec.~\ref{sec:fores} we
present the formulae for the computation of the electron spectrum. In
Sec.~\ref{sec:numev} we describe the numerical evaluation of the
spectrum. Section \ref{sec:disc} contains some discussion on the
different contributions in the high-energy region of the spectrum, and
the approximations we have used. We conclude in
Sec.~\ref{sec:concl}, where brief comments regarding the implications
of our results are given. Appendix \ref{app:conv} details the conventions
we use for the Dirac equation, and the electron and muon wavefunctions.

\section{Formulae for the electron spectrum}\label{sec:fores}
The Fermi interaction that
mediates muon decay is given by
\begin{equation}
\mathcal{L}_F=-2\sqrt{2}G_F\left[\bar{\psi}_{\nu_\mu}\gamma^{\rho}P_L\psi_{\mu}\right]\left[\bar{\psi}_{e}\gamma_{\rho}P_L\psi_{\nu_e}\right]+\textrm{h.c.},
\end{equation}
where $G_F=1.1663788(7)\times 10^{-5}\,\textrm{GeV}^{-2}$ is the Fermi constant and $P_L=(1-\gamma_5)/2$. This Lagrangian
can be Fierz rearranged to charge-retention ordering,
\begin{equation}\label{eq:LFfra}
\mathcal{L}_F=2\sqrt{2}G_F\left[\bar{\psi}_{e}\gamma^{\rho}P_L\psi_{\mu}\right]\left[\bar{\psi}_{\nu_\mu}\gamma_{\rho}P_L\psi_{\nu_e}\right]+\textrm{h.c.},
\end{equation}
which is the form that we will use. Since quantum electrodynamics
(QED) interactions do not affect the neutrino part of the Lagrangian, it is convenient
to partition the phase space and integrate the neutrino portion. In
that way, we generate an effective $\mu-e$ current and the free muon decay rate can be written as 
\begin{equation}
\Gamma =  \frac{1}{2E_{\mu}}\int dq^{2}\int\left[d\Pi_{\mu\to
    eq}\right]\left|\mathcal{M}_{\mu\to
    e}^{\rho\sigma}\right|^{2}T_{\rho\sigma},\label{eq:Gammfm}
\end{equation}
with
\begin{eqnarray}
d\Pi_{\mu\to eq} & \equiv &
\int\frac{d^{3}p_{e}}{(2\pi)^{3}2E_{e}}\int\frac{d^{3}k}{(2\pi)^{3}2E_{q}}(2\pi)^{4}\delta^{(4)}(p_{\mu}-p_{e}-q),
\nonumber \\
T_{\rho\sigma} & \equiv & -\frac{\pi}{3(2\pi)^{3}}\left(q^{2}g_{\rho\sigma}-q_{\rho}q_{\sigma}\right),
\nonumber \\
\left|\mathcal{M}_{\mu\to e}^{\rho\sigma}\right|^{2} & \equiv &  \frac{1}{2}\sum_{\mu^{-}\,\textrm{\tiny spin}}\,\sum_{e^{-}\,\textrm{\tiny spin}}8G_{F}^{2}\bar{u}(p_{e})\gamma^{\rho}P_{L}u(p_{\mu})\bar{u}(p_{\mu})\gamma^{\sigma}P_{L}u(p_{e}),
\end{eqnarray}
where the spinors $u(p)$ in that expression are normalized according
to $\bar{u}^{r}(p)u^{s}(p)=2m\delta^{rs}$, $q_{\rho}\equiv(E_{q},\vec{k})$
is the 4-momentum transferred to the neutrinos, $\vec{p}_{e}$ is
the electron three-momentum and $E_{e}$ and $E_{\mu}$
are the electron and muon energies, respectively. When we consider
the bound muon decay case, Eq.~(\ref{eq:Gammfm}) gets replaced by
\begin{eqnarray}
\Gamma &=&
\frac{2G_{F}^{2}}{(2\pi)^{6}}\sum_{e^{-}\,\textrm{\tiny spin}}\int dq^{2}\frac{d^{3}p_{e}}{E_{e}^{2}}\frac{d^{3}k}{E_{q}}(2\pi)\delta(E_{\mu}-E_{e}-E_{q})
\nonumber
\\
&&\times \left[\int d^{3}re^{-i\vec{k}\cdot\vec{r}}\bar{\varphi}_{e}\gamma^{\rho}P_{L}\varphi_{\mu}\right]\left[\int d^{3}r'e^{i\vec{k}\cdot\vec{r}'}\bar{\varphi}_{\mu}\gamma^{\sigma}P_{L}\varphi_{e}\right]T_{\rho\sigma},\label{eq:Gammbs}
\end{eqnarray}
where we are taking the nucleus as static (we discuss the inclusion
of recoil effects in the next Section), and $\varphi_{e}$
and $\varphi_{\mu}$ represent
the solutions of the Dirac equation for the electron and the muon,
respectively. We incorporate the average over the muon spin in the
definition of $\varphi_{\mu}$,
while we do not incorporate the sum over the electron spin in the
definition of $\varphi_{e}$. For the normalization
convention that is implied for the wavefunctions in
Eq.~(\ref{eq:Gammbs}) we refer to Appendix \ref{app:conv}. The muon energy in
Eq.~(\ref{eq:Gammbs}) is given by $E_{\mu}=m_{\mu}-E_{\mathrm{b}}$.
When the muonic atom is formed, the muon cascades down almost immediately
to the ground state, the $1S$ wavefunction should therefore be used
for $\varphi_{\mu}$ in
Eq.~(\ref{eq:Gammbs}) (the cascade process also depolarizes the
muons \cite{:2009yn}). We take the electron to be massless, since
electron mass effects are only relevant for $E_{e}\sim m_{e}$, which
is not our region of interest%
\footnote{In the low-energy region of the spectrum one should also consider
the possibility that the electron remains bound or captured by the nucleus.%
}. Integrating over $dq^{2}$ in Eq.~(\ref{eq:Gammbs}) we obtain
\begin{equation}
\Gamma=-\sum_{e^{-}\,\textrm{\tiny spin}}\frac{G_{F}^{2}}{192\pi^{7}}\int\frac{d^{3}p_{e}}{E_{e}^{2}}\int d^{3}kJ^{\rho}J^{\sigma\dagger}\left(q^{2}g_{\rho\sigma}-q_{\rho}q_{\sigma}\right),\label{eq:Gammuc}
\end{equation}
where it is understood that $q_{\rho}=(E_{\mu}-E_{e},\vec{k})$,
and we defined 
\begin{equation}
J^{\rho}\equiv\int d^{3}re^{-i\vec{k}\cdot\vec{r}}\bar{\varphi}_{e}\gamma^{\rho}P_{L}\varphi_{\mu}.\label{eq:defJ}
\end{equation}
The condition $q^{2}>0$ determines the limit of integration for $|\vec{k}|$
to be $|\vec{k}|<E_{\mu}-E_{e}$. Performing the angular
integration over $\vec{r}$ in the currents $J^{\rho}$, and the angular
integrations over $\vec{k}$ and $\vec{p}_e$ and summing over electron
spins in Eq.~(\ref{eq:Gammuc}) we obtain
\begin{eqnarray}\label{eq:espec}
\frac{1}{\Gamma_0}\frac{d\Gamma}{dE_e} & = &
\sum_{K\kappa}\frac{4}{\pi m_{\mu}^5}(2j_{\kappa}+1)\int_0^{E_{\mu}-E_e}\!\!\!\!\!\!\!\!\!\!dk\,
k^2
\nonumber\\
&&
\times \Bigg\{\left[(E_{\mu}-E_e)^2-k^2\right]\left(\frac{|S_{K\kappa}^{0}|^2}{K(K+1)}+\frac{|S_{K\kappa}^{-1}|^2}{K(2K+1)}+\frac{|S_{K\kappa}^{+1}|^2}{(K+1)(2K+1)}\right)
\nonumber\\
&& + \left[(E_{\mu}-E_e)k\right]2\textrm{Im}\!\!\left(\frac{S_{K\kappa}\left(S^{-1\,
      *}_{K\kappa}+S^{+1\,
      *}_{K\kappa}\right)}{2K+1}\right)+k^2\left(\frac{|S^{-1}_{K\kappa}+S^{+1}_{K\kappa}|^2}{(2K+1)^2}+|S_{K\kappa}|^2\right)\!\Bigg\},
\end{eqnarray}
where
\begin{equation}
\Gamma_{0}\equiv\frac{G_{F}^{2}m_{\mu}^{5}}{192\pi^{3}},
\end{equation}
is the free-muon decay rate. The $S$ functions in Eq.~(\ref{eq:espec}) are defined using
the notation $\left\langle \ldots\right\rangle \equiv\int_{0}^{\infty}\ldots r^{2}dr$,
and the two cases refer to odd/even $l_{\kappa}+K$, respectively
\begin{eqnarray}
S_{K\kappa}^{0} & = & \begin{cases}
-i\left(\kappa-1\right)\left\langle j_{K}(kr)\left(f_{\kappa}G+g_{\kappa}F\right)\right\rangle \\
\left(\kappa+1\right)\left\langle j_{K}(kr)\left(g_{\kappa}G-f_{\kappa}F\right)\right\rangle 
\end{cases}\nonumber ,\\
S_{K\kappa}^{-1} & = & \begin{cases}
\left\langle j_{K-1}(kr)\left[(\kappa-K-1)g_{\kappa}G-(\kappa+K-1)f_{\kappa}F\right]\right\rangle \\
-i\left\langle j_{K-1}(kr)\left[(\kappa+K+1)f_{\kappa}G+(\kappa-K+1)g_{\kappa}F\right]\right\rangle 
\end{cases}\nonumber ,\\
S_{K\kappa}^{+1} & = & \begin{cases}
\left\langle j_{K+1}(kr)\left[(\kappa+K)g_{\kappa}G+(K-\kappa+2)f_{\kappa}F\right]\right\rangle \\
-i\left\langle j_{K+1}(kr)\left[(\kappa-K)f_{\kappa}G+(\kappa+K+2)g_{\kappa}F\right]\right\rangle 
\end{cases}\nonumber ,\\
S_{K\kappa} & = & \begin{cases}
i\left\langle j_{K}(kr)(f_{\kappa}G-g_{\kappa}F)\right\rangle \\
\left\langle j_{K}(kr)\left(g_{\kappa}G+f_{\kappa}F\right)\right\rangle 
\end{cases},\label{eq:functionsS}
\end{eqnarray}
where $G$ and $F$ ($g_{\kappa}$ and $f_{\kappa}$)
are the upper and lower components of the radial muon (electron) wavefunction,
respectively. $\kappa$ is the quantum number appearing in the Dirac equation (see Appendix \ref{app:conv} for the conventions we use and the definitions of $j_{\kappa}$ and
$l_{\kappa}$). For a given value of $K$ in Eq.~(\ref{eq:espec}), $\kappa$ can only take the
values $\pm K$ and $\pm(K+1)$. The sum over $K$ goes from 0 to
$\infty$, but $K$ cannot take the value $K=0$ in the $S^0_{K\kappa}$ and
$S^{-1}_{K \kappa}$ terms, and $\kappa$ can never be equal to 0. $j_n(z)$ is the
spherical Bessel function of order $n$. Equation (\ref{eq:espec})
agrees with the expressions presented in Refs.~\cite{Watanabe:1987su,Watanabe:1993}.

\subsection{Inclusion of recoil effects}\label{subsec:recoil}
In the previous Section we considered the nucleus to be static.
The upcoming conversion experiments plan to use an aluminum target
(previous conversion experiments used heavier elements), where the atomic
mass of aluminum ($Z=13$, $A=27$) is $25133\,\textrm{MeV}$. The nucleus is, therefore,
more than 200 times heavier than the muon ($m_{\mu}=105.6584\,\textrm{MeV}$)
and recoil effects should be negligible for most of the electron
spectrum. However, the nuclear-recoil energy modifies the endpoint of the electron
spectrum $E_{\mu e}$, see Eq.~(\ref{eq:conven}),
which means that recoil effects need to be carefully considered in
studies of the
high-energy part of the spectrum. We will always consider the recoil
effects at first order in a $1/m_{N}$ expansion, where $m_{N}$ is the mass of
the nucleus $N$. 

For muon DIO, the nuclear-recoil energy is 
\begin{equation}
E_{\mathrm{rec}}=\frac{|\vec{p}_{N}|^{2}}{2m_{N}},
\end{equation}
where the three-momentum of the nucleus is
\begin{equation}
\vec{p}_{N}=-\vec{p}_{e}-\vec{p}_{\bar{\nu}_{e}}-\vec{p}_{\nu_{\mu}},
\end{equation}
with $\vec{p}_{\nu_{\mu}}$
and $\vec{p}_{\bar{\nu}_{e}}$ the three-momenta
of the neutrino and anti-neutrino, respectively. We see that, for a given electron
energy, the nuclear-recoil energy is not constant but depends on the
momenta of the neutrinos. This complicates the integration over the
neutrino momenta, but in the high-energy end of the spectrum we can
approximate 
\begin{equation}
E_{\mathrm{rec}}=\frac{\left|\vec{p}_{N}\right|^{2}}{2m_{N}}=\frac{\left(\vec{p}_{e}+\vec{p}_{\bar{\nu}_{e}}+\vec{p}_{\nu_{\mu}}\right)^{2}}{2m_{N}}\simeq\frac{\left|\vec{p}_{e}\right|^{2}}{2m_{N}}=\frac{E_{e}^{2}}{2m_{N}},\label{eq:Erecapp}
\end{equation}
so that the recoil effects amount to a change in the momentum transfer
to the neutrinos. The net effect is to substitute
\begin{equation}
E_{\mu}-E_e\to E_{\mu}-E_e-\frac{E_e^2}{2m_N},
\end{equation}
in the upper limit of the integration over $k$ and inside the square
brackets in
Eq.~(\ref{eq:espec}).
The endpoint of the electron spectrum
is given by 
\begin{equation}
E_{\mu e}=E_{\mu}-\frac{E_{\mu}^{2}}{2m_{N}}=m_{\mu}-E_{\mathrm b}-\frac{E_{\mu}^{2}}{2m_{N}},
\end{equation}
which is exact up to corrections of order $1/m_{N}^{2}$.
The approximation
for the recoil energy in Eq.~(\ref{eq:Erecapp}) is the same as used in Ref.~\cite{Shanker:1981mi}. The electron spectrum including nuclear-recoil effects is, therefore,
given by
\begin{eqnarray}
\frac{1}{\Gamma_{0}}\frac{d\Gamma}{dE_{e}} & = &
\sum_{K\kappa}\frac{4}{\pi m_{\mu}^{5}}(2j_{\kappa}+1)\int_{0}^{E_{\mu}-E_{e}-\frac{E_{e}^{2}}{2m_{N}}}\!\!\!\!\!\!\!\!\!\! dk\, k^{2}
\nonumber \\
 &  &
\times
 \Bigg\{\left[\left(E_{\mu}-E_{e}-\frac{E_{e}^{2}}{2m_{N}}\right)^{2}\!\!\!-k^{2}\right]\!\left(\frac{|S_{K\kappa}^{0}|^{2}}{K(K+1)}+\frac{|S_{K\kappa}^{-1}|^{2}}{K(2K+1)}+\frac{|S_{K\kappa}^{+1}|^{2}}{(K+1)(2K+1)}\right)
 \nonumber \\
 &  & + 2k\left(E_{\mu}-E_{e}-\frac{E_{e}^{2}}{2m_{N}}\right)\textrm{Im}\frac{S_{K\kappa}\left(S_{K\kappa}^{-1\,*}+S_{K\kappa}^{+1\,*}\right)}{2K+1}
 \nonumber \\
 &  & + k^{2}\left(\frac{|S_{K\kappa}^{-1}+S_{K\kappa}^{+1}|^{2}}{(2K+1)^{2}}+|S_{K\kappa}|^{2}\right)\!\Bigg\},\label{eq:especrec}
\end{eqnarray}
where it is understood that this expression
should only be used in the region where Eq.~(\ref{eq:Erecapp}) is
a good approximation to the nuclear-recoil energy. As will be manifest
in the following Sections, recoil effects become negligible
before Eq.~(\ref{eq:Erecapp}) ceases to be a good approximation to
the recoil energy. That means the inclusion of recoil effects
beyond the approximation considered here is unnecessary. 

\subsection{Endpoint expansions}
Equation (\ref{eq:especrec}) constitutes our final result for the
high-energy region of the electron spectrum and it is what we will
use in our numerical evaluations. Nevertheless, it is still interesting
to perform a Taylor expansion of Eq.~(\ref{eq:especrec}) around
the endpoint, to make the behavior of the spectrum manifest. We obtain
\begin{eqnarray}
\left.\frac{1}{\Gamma_{0}}\frac{d\Gamma}{dE_{e}}\right\vert
_{E_{e}\sim E_{\mu}-\frac{E_{\mu}^{2}}{2m_{N}}} & = & \frac{64}{5\pi
  m_{\mu}^{5}}\left(E_{\mu}-E_{e}-\frac{E_{e}^{2}}{2m_{N}}\right)^{5}\left(p_{1}^{2}+\frac{s_{1}^{2}}{3}+\frac{2}{3}r_{2}^{2}\right)
\nonumber \\
&\equiv & 
\label{eq:TErec}
B \left(E_{\mu}-E_{e}-\frac{E_{e}^{2}}{2m_{N}}\right)^{5},
\end{eqnarray}
where $p_{\kappa}=\left\langle g_{-\kappa}G\right\rangle $,
$s_{\kappa}=\left\langle f_{-\kappa}F\right\rangle ,$
$r_{\kappa}=\left\langle g_{-\kappa}F\right\rangle $,
and it is understood that the electron wavefunctions $g_{\kappa}$
and $f_{\kappa}$ in Eq.~(\ref{eq:TErec}) correspond
to the energy $E_{e}=E_{\mu}-E_{\mu}^{2}/(2m_{N})$. We show the values
of the $B$ coefficient in Eq.~(\ref{eq:TErec}), for a few elements, in
Table \ref{tab:Bcoef}.
\begin{table}
\caption{Values for the $B$ coefficient of the leading-order Taylor
  expansion in Eq.~(\ref{eq:TErec}), for a few elements. We use
  finite-size nuclei, characterized by a two-parameter Fermi
  distribution (see Eq.~(\ref{eq:2pFd})), with the values of the
  parameters of that distribution taken from Refs.~\cite{De Jager:1987qc,Fricke:1995zz}.}\label{tab:Bcoef}
\begin{ruledtabular}
\begin{tabular}{cc}
Nucleus & $B(\,\textrm{MeV}^{-6})$\\
\hline
Al($Z=13$) & $8.98\times 10^{-17}$ \\
Ti($Z=22$) & $4.94\times 10^{-16}$ \\
Cu($Z=29$) & $1.14\times 10^{-15}$ \\
Se($Z=34$) & $1.62\times 10^{-15}$ \\
Sb($Z=51$) & $3.57\times 10^{-15}$ \\
Au($Z=79$) & $4.79\times 10^{-15}$ \\
\end{tabular}
\end{ruledtabular}
\end{table}
The corresponding Taylor expansion for the
expression without recoil effects in Eq.~(\ref{eq:espec}) is given by
\begin{equation}\label{eq:TEnorec}
\left. \frac{1}{\Gamma_0}\frac{d\Gamma}{dE_e}\right\vert_{E_e\sim
  E_{\mu}}^{\textrm{\tiny no recoil}}=\frac{64}{5\pi m_{\mu}^5}(E_{\mu}-E_e)^5\left(p_1^2+\frac{s_1^2}{3}+\frac{2}{3}r_2^2\right),
\end{equation}
where it is understood that the electron wavefunctions in this
equation correspond to the energy
$E_e=E_{\mu}$. Our Taylor expansion agrees with the results in Ref.~\cite{Shanker:1981mi}.

\section{Numerical evaluation of the spectrum}\label{sec:numev}

We now use Eq.(\ref{eq:especrec}) to obtain a numerical evaluation
of the high-energy region of the electron spectrum. We present the
results for the case of an aluminum nucleus (Al, $Z=13$), which is
the target intended to be used in Mu2e and COMET
\cite{Carey:2008zz,Cui:2009zz}. 

We consider a nucleus of finite size, characterized by a two-parameter
Fermi distribution $\rho(r)$, given by 
\begin{equation}
\rho(r)=\rho_{0}\frac{1}{1+e^{\frac{r-r_{0}}{a}}}.\label{eq:2pFd}
\end{equation}
 For the parameters of the Fermi distribution we use the values \cite{De Jager:1987qc}
\begin{equation}\label{eq:parFd}
r_0=2.84\pm 0.05\,\textrm{fm},\quad a=0.569\,\textrm{fm}.
\end{equation}
$\rho_0$ in Eq.~(\ref{eq:2pFd}) is the normalization factor, which can be
expressed in terms of $r_{0}$ and $a$. For the muon mass, aluminum
mass and the fine-structure constant we use the values $m_{\mu}=105.6584\,\textrm{MeV},\text{ }m_{\textrm{\tiny Al}}=25133\,\textrm{MeV},\text{ }\alpha=\frac{1}{137.036},$
and remember that we take the electron to be massless. We numerically
solve the radial Dirac equations for the muon and the electron, with
the charge distribution in Eq.~(\ref{eq:2pFd}), to obtain the wavefunctions.
For the muon energy we obtain 
\begin{equation}
E_{\mu}=m_{\mu}-E_{\textrm b}=105.194\,\textrm{MeV},
\end{equation}
which gives the endpoint energy 
\begin{equation}\label{eq:Emue}
E_{\mu e}=E_{\mu}-\frac{E_{\mu}^{2}}{2m_{\textrm{\tiny
      Al}}}=104.973\,\textrm{MeV}.
\end{equation}
Electron screening will increase the end-point energy in
Eq.~(\ref{eq:Emue}) by about $+0.001\,{\textrm MeV}$ and similarly
shift the overall spectrum. That small effect is negligible for our
considerations. Recall that the sum over $K$ in Eq.~(\ref{eq:especrec}) goes from 0 to $\infty$,
we include as many terms in $K$ as necessary in order to get three-digit precision for each point of the spectrum. This
requires about 30 terms near $m_{\mu}/2$ and fewer terms in the low-
and the high-energy parts of the spectrum. 

\begin{figure}
\centering
\includegraphics[width=8cm]{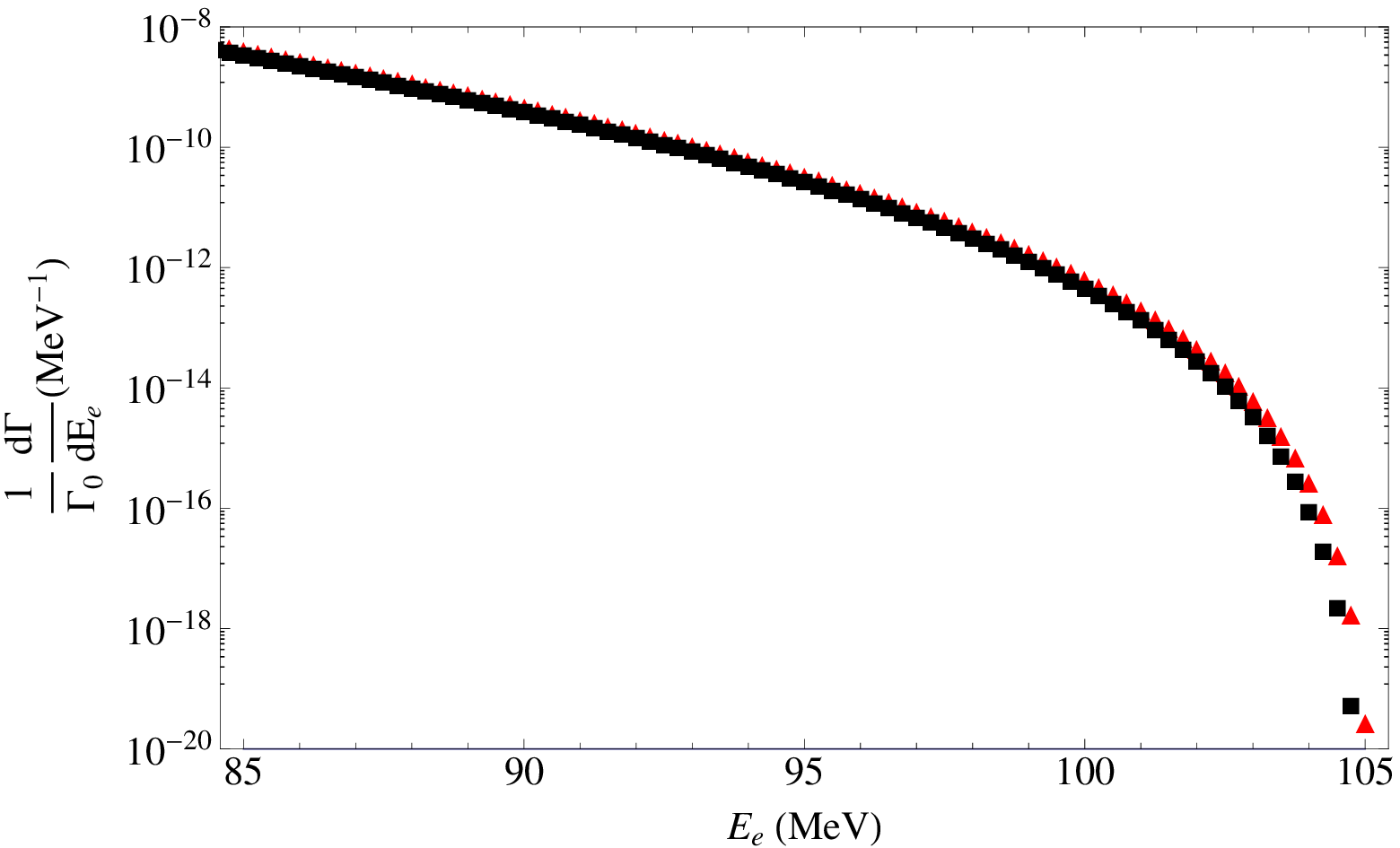}\hspace{.4cm}\includegraphics[width=8cm]{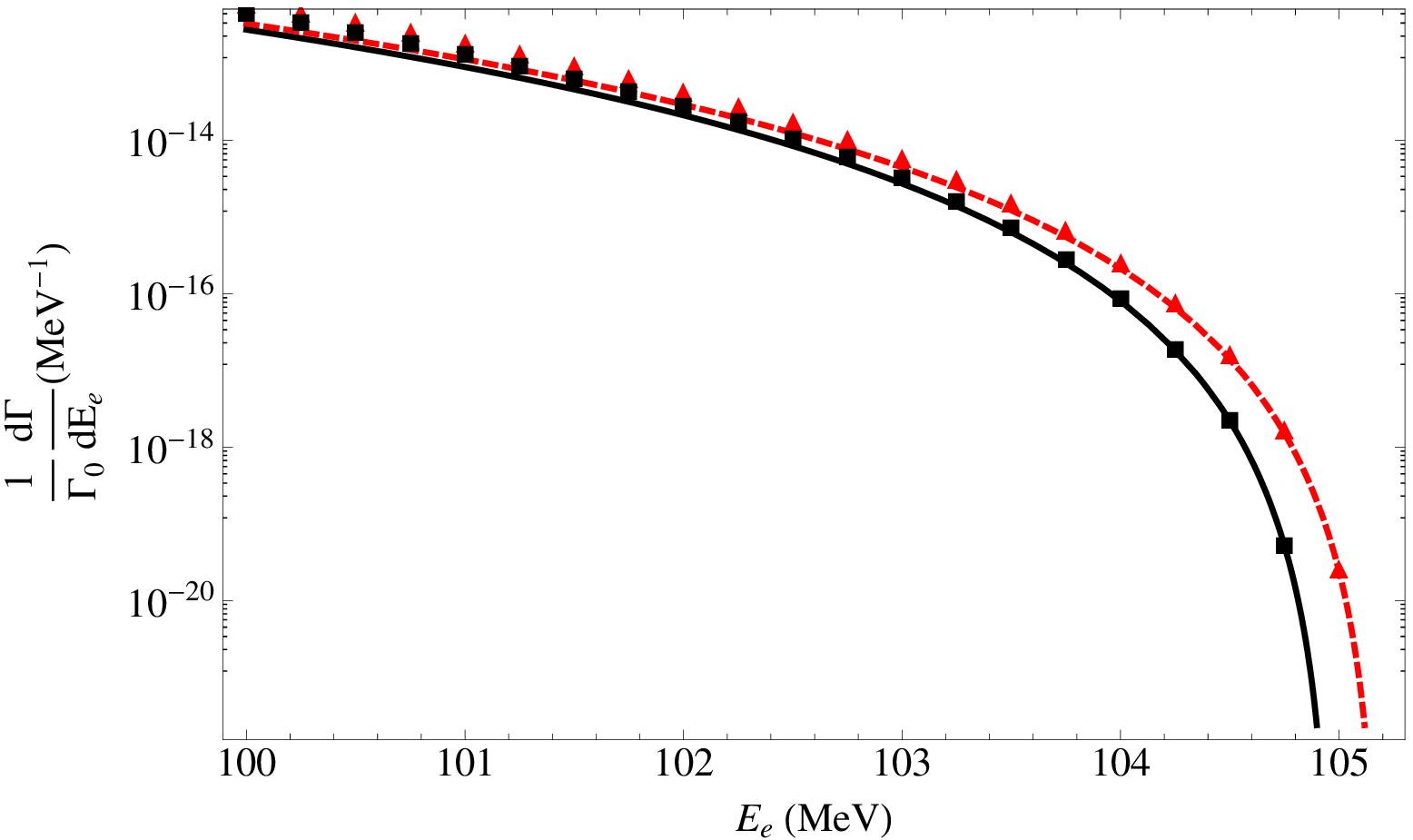}
\caption{Endpoint region of the electron spectrum for aluminum. The squares correspond to the spectrum with recoil effects,
  Eq.~(\ref{eq:especrec}). For comparison, we show the spectrum
  neglecting recoil, Eq.~(\ref{eq:espec}), as the triangles. The right plot is
  a zoom for $E_e>100\,\textrm{MeV}$, the solid (dashed) line on this plot
  corresponds to the Taylor expansion around the endpoint with
  (without) recoil. 
}\label{fig:speceprec}
\end{figure}

\begin{figure}
\centering \includegraphics[width=8cm]
{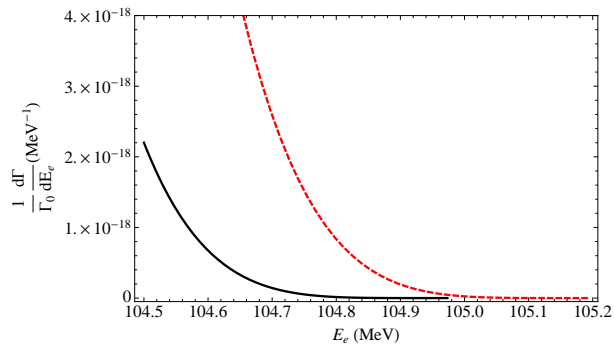}
\caption{Detail of the electron spectrum for aluminum very close to the high-energy
endpoint with (neglecting) nuclear recoil, represented  as the solid (dashed) line.}
\label{fig:speceplsc} 
\end{figure}

We present the result of the numerical evaluation of the high-energy
region of the electron spectrum in Fig~\ref{fig:speceprec}. The squares in the figure are the spectrum with recoil effects,
from Eq.~(\ref{eq:especrec}). For comparison, we also show the result
obtained by neglecting recoil effects, from Eq.~(\ref{eq:espec}), as the triangles.
The right plot in the figure is a zoom for $E_{e}>100\,\textrm{MeV}$,
the solid and dashed lines on this plot correspond to the
Taylor expansions in Eqs.~(\ref{eq:TErec}) and (\ref{eq:TEnorec}),
respectively. 
Terms up to $K=4$ were included in Fig.~\ref{fig:speceprec}.
Figure \ref{fig:speceplsc} presents a detail of the electron spectrum
very close to the high-energy endpoint in linear scale. We can appreciate
in that figure how the spectra with (solid line) and without
(dashed line) recoil effects tend to zero at the corresponding
endpoints (the endpoint without recoil is at $E_{e}=E_{\mu}$).
To make our results easier to use, we mention that the polynomial
\begin{equation}\label{eq:fitep}
P(E_e)\equiv a_5\delta^5+a_6\delta^6+a_7\delta^7+a_8\delta^8,
\end{equation}
with
\begin{equation}
a_5=8.6434\times 10^{-17},\, a_6=1.16874\times 10^{-17},\,
a_7=-1.87828\times 10^{-19},\, a_8=9.16327\times 10^{-20},
\end{equation}
the energies expressed in MeV, and 
\begin{equation}
\delta=E_{\mu}-E_e-\frac{E_e^2}{2m_{\textrm{\tiny
      Al}}},
\end{equation}
fits very well the result for the electron spectrum in aluminum
normalized to the free decay rate (squares in
Fig.~\ref{fig:speceprec}) for all $E_e>85\,\textrm{MeV}$ (i.e., the difference between Eq.~(\ref{eq:fitep}) and
the squares in Fig.~\ref{fig:speceprec} is not larger than the uncertainties discussed in the next Section). Note that, in order to obtain a better fit for the whole $E_e>85$ MeV region, the value of $a_5$, in Eq.~(\ref{eq:fitep}), was not constrained to be
that of the leading coefficient of the Taylor expansion in Table
\ref{tab:Bcoef}.

\begin{figure}
\centering
\includegraphics[width=8cm]{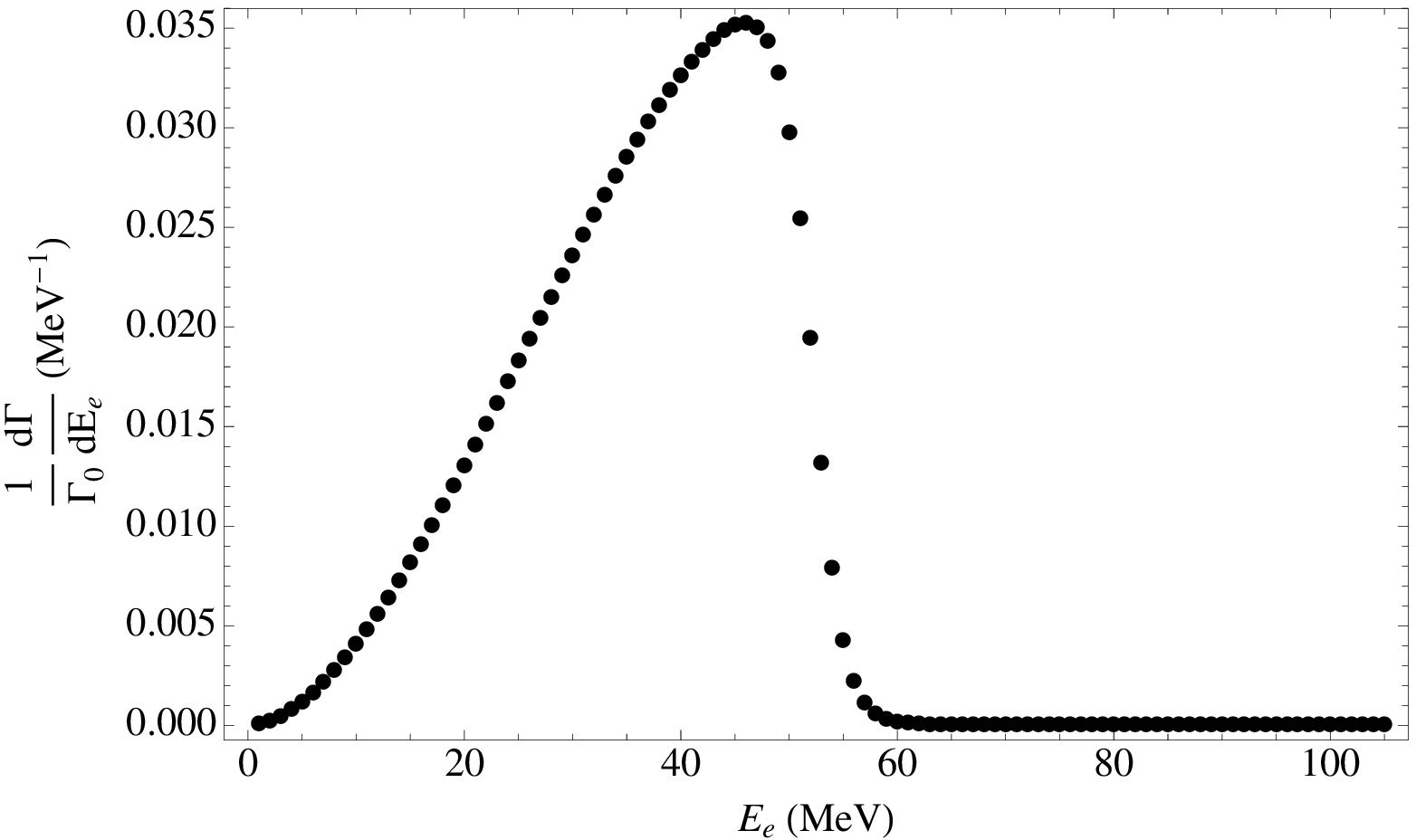}\hspace{.4cm}\includegraphics[width=8cm]{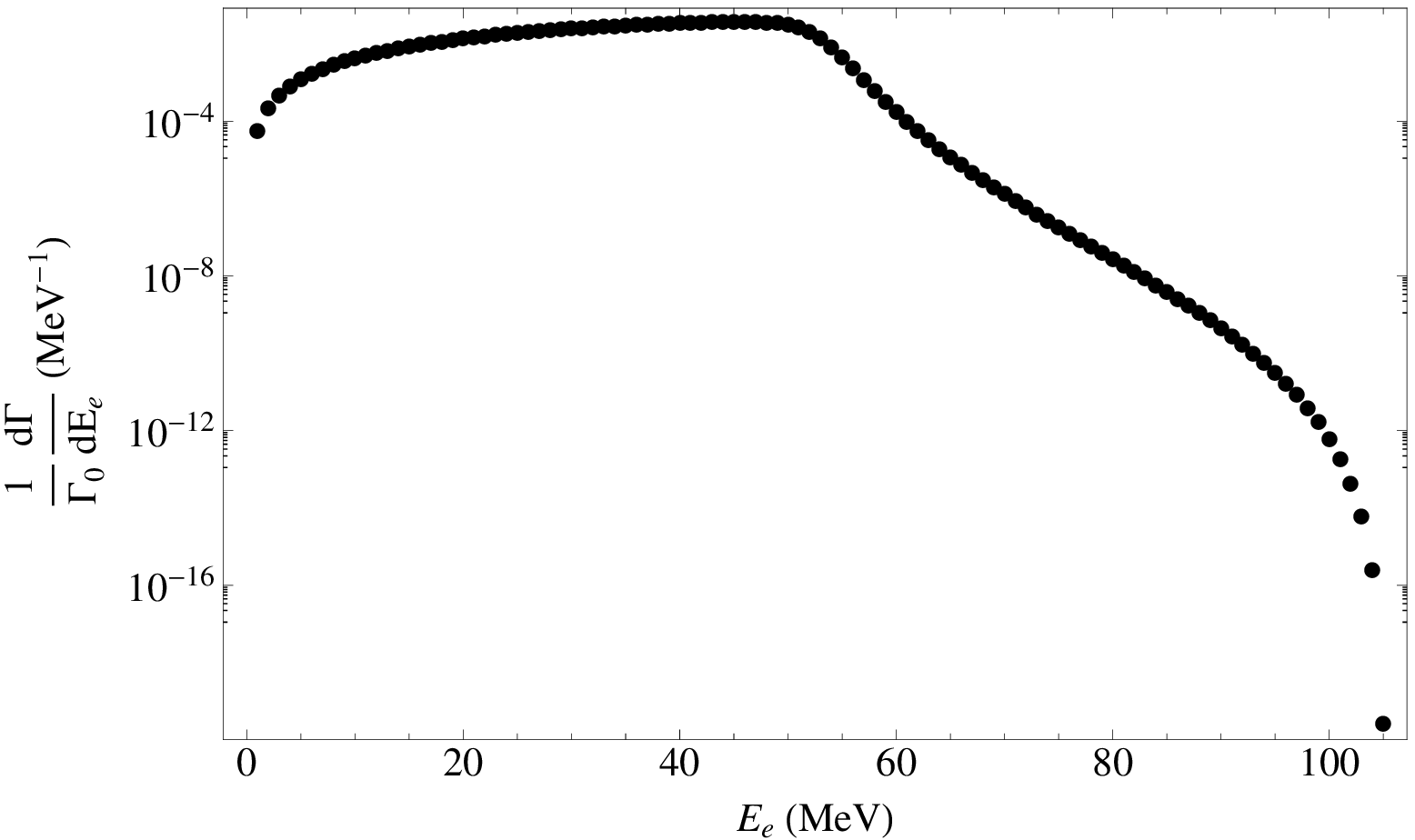}
\caption{Electron spectrum for aluminum. Left plot: linear scale;
  right plot: logarithmic scale.}\label{fig:spec}
\end{figure}

For completeness, we also show the spectrum for the full range of
electron energies in Fig.~\ref{fig:spec} as the circles, from
Eq.~(\ref{eq:espec}). Terms up to $K=31$ were included in this plot. 
The total decay rate for muon decay in orbit in aluminum is obtained by
integrating the spectrum in Fig.~\ref{fig:spec}. The result we
obtain is
\begin{equation}
\frac{1}{\Gamma_0}\int_0^{E_{\mu e}}
\frac{d\Gamma}{dE_e} dE_e=0.9934,
\end{equation}
in agreement with Ref.~\cite{Watanabe:1993}. Nuclear-recoil effects are negligible in the total rate.  Also note, the integrated ordinary muon decay rate
is hardly affected by the presence of the Coulomb potential, in accord
with the general results in \cite{Uberall:1960zz,Czarnecki:1999yj}.

Since the Mu2e Collaboration also considers titanium (Ti, $Z=22$) as a
viable target \cite{Carey:2008zz}, we give the polynomial, $P^{(\textrm{\tiny
      Ti})}(E_e)$, that fits
the result for the electron spectrum in titanium (normalized to the
free decay rate) for energies $E_e>85\,\textrm{MeV}$,
\begin{equation}\label{eq:fitepTi}
P^{(\textrm{\tiny
      Ti})}(E_e)\equiv a_5^{(\textrm{\tiny
      Ti})}\delta_{(\textrm{\tiny
      Ti})}^{ 5}+a_6^{(\textrm{\tiny
      Ti})}\delta_{(\textrm{\tiny
      Ti})}^6+a_7^{(\textrm{\tiny
      Ti})}\delta_{(\textrm{\tiny
      Ti})}^7+a_8^{(\textrm{\tiny
      Ti})}\delta_{(\textrm{\tiny
      Ti})}^8,
\end{equation}
with
\begin{equation}
a_5^{(\textrm{\tiny
      Ti})}=4.44278\times 10^{-16},\, a_6^{(\textrm{\tiny
      Ti})}=9.06648\times 10^{-17},\,
a_7^{(\textrm{\tiny
      Ti})}=-4.26245\times 10^{-18},\, a_8^{(\textrm{\tiny
      Ti})}=8.193\times 10^{-19},
\end{equation}
the energies expressed in MeV,
\begin{equation}
\delta_{(\textrm{\tiny
      Ti})}=E_{\mu}-E_e-\frac{E_e^2}{2m_{\textrm{\tiny
      Ti}}},
\end{equation}
and for titanium $E_{\mu}=104.394\,\textrm{MeV}$.

\section{Discussion}\label{sec:disc}
As expected, the results in Figure~\ref{fig:speceprec} show that
nuclear-recoil effects are only important close to the high-energy
endpoint of the spectrum. The corrections to the approximation of
the recoil energy that we have used in Eq.~(\ref{eq:Erecapp}) are
of order $E_{e}(E_{\mu}-E_{e})/(2m_{N})$ (while Ref.~\cite{Shanker:1981mi} seems to wrongly estimate
this correction as being smaller $\sim\left(E_{\mu}-E_{e}\right)^{2}/(2m_{N})$).
For electron energies around $85\,\textrm{MeV}$, Eq.~(\ref{eq:Erecapp})
is still a good approximation to the recoil energy while the effect
of recoil in the spectrum is very small. When the corrections
to the approximation in Eq.~(\ref{eq:Erecapp}) become order one,
the recoil effects on the spectrum are negligible. Therefore we conclude
that, as anticipated in Sec.~\ref{subsec:recoil}, inclusion of recoil
effects beyond the approximation considered here is unnecessary.

As already noted in Ref.~\cite{Shanker:1981mi}, the Schr\"odinger
wave function for the muon is not a good approximation near the
endpoint. In that region, one needs to produce an electron with $E_{e}\sim|\vec{p}_{e}|\sim m_{\mu}$.
This implies that, either the muon has $|\vec{p}_{\mu}|\sim m_{\mu}$
(i.e., it is at the tail of the wavefunction) or (if the muon has
the typical atomic non-relativistic momentum, of order the inverse
Bohr radius) the electron must interact with the nucleus to get
$|\vec{p}_{e}|\sim m_{\mu}$. Those two contributions
are of the same order in $\alpha$, which means that we cannot
treat the muon within a non-relativistic approximation. There are,
thus, some leading contributions where the muon is far off-shell (it has
$E_{\mu}\sim|\vec{p}_{\mu}|\sim m_{\mu}$),
a fact that also tells us that this is the region where finite-nuclear-size
effects will be most important (since the muon will be closer to the
nucleus). By this argument, we can also understand that only the lowest
values of the angular momentum in the electron wavefunctions contribute
at the endpoint (as we see in Eq.~(\ref{eq:TErec})).

Uncertainties in the modeling of finite nuclear-size effects
can induce errors in the electron spectrum and, as we discussed in the
previous paragraph, those are expected to be most important in the
endpoint region. We re-computed the spectrum varying the parameters in
the Fermi distribution as indicated in Eq.~(\ref{eq:parFd}), and found
that the errors induced in the spectrum do increase as we approach
the endpoint, as expected, but they are never larger than $\pm2\%$;
so, we
can safely ignore them.

Finally we comment that  radiative corrections have not been included in Eq.~(\ref{eq:especrec}).  However, they are
not expected to significantly modify the results presented.

\section{Conclusions and Experimental Applications}\label{sec:concl}

\begin{figure}
\centering\includegraphics[width=8cm]{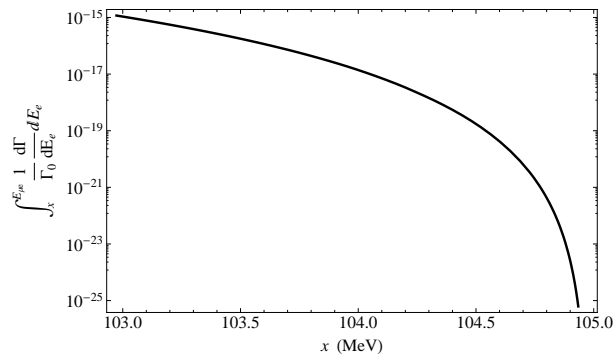}
\caption{Total rate of decay-in-orbit events, for aluminum, with electron energy larger
than $x$, normalized to the free muon decay rate $\Gamma_{0}$.\label{fig:intxep}}
\end{figure}

We have performed a detailed evaluation of the high-energy region
of the electron spectrum in muon decay in orbit. Our results in
Eq.~(\ref{eq:especrec}), Eq.~(\ref{eq:fitep})
and Figure~\ref{fig:speceprec} include all the relevant effects
to accurately describe the high-energy region of the spectrum, and
provide the correct background contribution for $\mu-e$
conversion search experiments. To summarize our findings, we plot in
Fig.~\ref{fig:intxep} the rate of decay in orbit events producing an
electron with energy higher than $x$, as a function of $x$ (normalized to the
free muon decay rate).

The complete muon DIO electron spectrum presented here provides a
check on previous low- and high-energy partial calculations \cite{Watanabe:1993,Shanker:1981mi}, as well as
an interpolation between them. It properly incorporates recoil and
relativistic effects in the high-energy endpoint region, which  is of
crucial importance for future $\mu-e$ conversion background
studies. In that regard, its primary utility is twofold. First, when
experimental data on muon DIO becomes available, our formula in
Eq.~(\ref{eq:fitep})  can be compared with it and used to refine the
detector's acceptance, efficiency and resolution. Second, the expected
spectrum can be convoluted with the spectrometer resolution function
to obtain a more precise estimate of the muon DIO  background to CLFV
$\mu-e$ conversion in the $E_e \simeq 103.5-105$ MeV signal region.

Currently available estimates by the Mu2e Collaboration \cite{Carey:2008zz} find for
$4\times 10^{16}$  ordinary muon captures in Al (corresponding to a
total of $2.6 \times 10^{16}$ DIO) a signal of 4 conversion events if $R_{\mu \mathrm{Al}} \equiv 
\Gamma\left( \mu^- \mathrm{Al} \to e^- \mathrm{Al}\right) /
\Gamma\left( \mu^- \mathrm{Al} \to \nu_\mu \mathrm{Mg}\right)
=10^{-16}$, with only about 0.2 DIO background events for their detector
resolution function. At that level, the discovery capability is quite
robust. However, if $R_{\mu \mathrm{Al}}$  is much smaller, more running will be
required to enhance the signal and refinements of the background using
the spectrum in Eq.~(\ref{eq:fitep}) will be critical.

One can also use our analysis to make a rough comparison of DIO
backgrounds for stopping targets with different $Z$.  Indeed, if a
signal for $\mu\to e$  conversion is found in Al, other targets will be
important for extracting the underlying ``new physics'' responsible
for it. Of course, for much higher $Z$, the initial dead time of 700
ns envisioned in the Mu2e proposal for eliminating prompt backgrounds
would significantly reduce the number of ``live'' muon captures and
severely compromise the experimental sensitivity. Ignoring that issue
for now, we can ask: What DIO background is expected for a higher-$Z$
target with energy resolution identical to the Al setup of Mu2e, while
continuing to require that it also produces 4 signal events?
Considering the case of a titanium target, with $Z=22$, we expect $ 
R_{\mu \mathrm{Ti}} \simeq 1.6R_{\mu \mathrm{Al}}$ for models of ``new
physics'' \cite{Marciano:2008zz} dominated by chiral changing CLFV. So, one needs a ``live''
run with ``only'' $2.5 \times 10^{16}$ ordinary muon captures or
correspondingly $0.43\times 10^{16}$ total DIO events to reach the
same 4 event discovery sensitivity as in aluminum. However, even though
the total number of DIO events (for all $E_e$) is smaller by a factor
of 6 in Ti, the relative branching fraction for high-energy DIO
events in the signal region (with similar detector resolution) is
about 6 times larger for Ti compared to Al. So, overall the DIO
background is about the same for Ti. 

More difficult will be the loss of muon events in higher $Z$ materials
due to the 700 ns dead time during which most of the muons undergo
capture. For that, a complete reassessment of the muon production and
stopping conditions may be required. 

\begin{acknowledgments} This research was supported by Science and
Engineering Research Canada and by the United States Department of
Energy under Grant Contract DE-AC02-98CH10886. \end{acknowledgments}

\appendix

\section{Conventions}\label{app:conv}
In this Appendix we explain our conventions for the Dirac
equation, and the electron and muon wavefunctions.

The Dirac equation in a central field is given by
\begin{equation}
W\psi=\left[-i\gamma_5\sigma_r\left(\frac{\partial}{\partial r}+\frac{1}{r}-\frac{\beta}{r}K\right)+V(r)+m\beta\right]\psi,
\end{equation}
where $W$ is the energy of the particle, $m$ its mass and $V(r)$ is
the potential. The $4\times 4$ matrices $\gamma_5$, $\beta$, $K$ and $\sigma_r$ are given
by
\begin{equation}
\gamma_5=\left(\begin{array}{cc}0 & 1\\ 1 & 0\end{array}\right)\, ,\,
\beta=\left(\begin{array}{cc}1 & 0\\ 0 & -1\end{array}\right)\, , \,
r\sigma_r=\left(\begin{array}{cc}\vec{\sigma}\cdot \vec{r} & 0\\ 0
    & \vec{\sigma}\cdot \vec{r}\end{array}\right)\, , \,
K=\left(\begin{array}{cc}\vec{\sigma}\cdot \vec{l}+1 & 0\\ 0
    & -(\vec{\sigma}\cdot \vec{l}+1)\end{array}\right),
\end{equation}
with $\vec{l}$ the orbital angular momentum $\vec{l}=-i
\vec{r}\times \vec{\nabla}$ and $\vec{\sigma}$ the
$2\times 2$ Pauli matrices
\begin{equation}
\sigma_x=\left(\begin{array}{cc}0 & 1\\ 1 & 0\end{array}\right)\, ,\,
\sigma_y=\left(\begin{array}{cc}0 & -i\\ i & 0\end{array}\right)\, , \,
\sigma_z=\left(\begin{array}{cc}1 & 0\\ 0 & -1\end{array}\right).
\end{equation}
The wavefunctions are generically denoted as follows
\begin{equation}
\psi=\psi_{\kappa}^{\mu}=\left(\begin{array}{c}g_{\kappa}(r)\chi_{\kappa}^{\mu} \\ if_{\kappa}(r)\chi_{-\kappa}^{\mu}\end{array}\right),
\end{equation}
they diagonalize the operators $K$, $\vec{j}^2$ and $j_z$
($\vec{j}$ being the total angular momentum) with eigenvalues
$-\kappa$, $j(j+1)$ and $\mu$, respectively. $g_{\kappa}$ and $f_{\kappa}$ are the radial functions which are given by the equations
\begin{eqnarray}
\frac{df_{\kappa}}{dr} & = & \frac{\kappa-1}{r}f_{\kappa}-(W-m-V)g_{\kappa}, \\
\frac{dg_{\kappa}}{dr} & = & (W-V+m)f_{\kappa}-\frac{\kappa+1}{r}g_{\kappa}.
\end{eqnarray}
$\chi_{\kappa}^{\mu}=\chi_{\kappa}^{\mu}(\hat{r})$ are the spin-angular functions, which satisfy
\begin{equation}
(\vec{\sigma}\cdot \vec{l}+1)\chi_{\kappa}^{\mu}=-\kappa \chi_{\kappa}^{\mu}\, ,\, j_z \chi_{\kappa}^{\mu}=\mu
\chi_{\kappa}^{\mu}\, ,\,\int d\Omega_{\hat{r}}\chi_{\kappa}^{\mu\dagger}\chi_{\kappa'}^{\mu'}=\delta_{\mu\mu'}\delta_{\kappa\kappa'}.
\end{equation}
They are given by
\begin{equation}\label{eq:chikm}
\chi_{\kappa}^{\mu}=\sum_mC\left( l\frac{1}{2}j;\mu-m\, m\, \mu\right)Y_{l}^{\mu-m}\chi^m,
\end{equation}
with $C(lsj;l_zs_zj_z)$ the Clebsch-Gordan
coefficients, $Y_l^{\mu}$ the spherical harmonics and $\chi^m$ the spin
$1/2$ eigenfunctions
\begin{equation}
\vec{s}^{\, 2}\chi^m=\frac{3}{4}\chi^m\, ,\, s_z\chi^m=m\chi^m\, ,\, m=\pm\frac{1}{2},
\end{equation}
where $\vec{s}=\frac{1}{2} \vec{\sigma}$. Eq.~(\ref{eq:chikm}) makes
manifest that $\chi_{\kappa}^{\mu}$ is an eigenfunction of $\vec{\sigma}\cdot
\vec{l}+1=\vec{j}^{\, 2}-\vec{l}^{\, 2}-\vec{s}^{\, 2}+1$
with eigenvalue 
\begin{equation}
(\vec{\sigma}\cdot\vec{l}+1) \chi_{\kappa}^{\mu}=\left[j(j+1)-l(l+1)+\frac{1}{4}\right]\chi_{\kappa}^{\mu}\equiv-\kappa \chi_{\kappa}^{\mu}.
\end{equation}
Thus we have
\begin{equation}
\kappa=\left\{\begin{array}{cc}l & \textrm{for } j=l-\frac{1}{2}\\ -l-1 & \textrm{for } j=l+\frac{1}{2}\end{array}\right.,
\end{equation}
and we see that $\kappa$ can take all integer values except 0. We also note
that the value of $j$ is given by $\kappa$ according to
\begin{equation}\label{eq:jkappa}
j=|\kappa|-\frac{1}{2}\equiv j_{\kappa},
\end{equation}
and that the value of $l$ is also given by $\kappa$, according to
\begin{equation}
l=j+\frac{1}{2}\frac{\kappa}{|\kappa|}\equiv l_{\kappa}.
\end{equation}

We express the $1S$ muon wavefunction as
\begin{equation}
\varphi_{\mu}(\vec{r})=\sum_sa_s \left(\begin{array}{c}G\chi_{-1}^s\\iF\chi_{1}^s\end{array}\right),
\end{equation}
where $a_s$ is the amplitude of the muon state with spin
projection $s$, since we need an unpolarized muon we have
$|a_s|^2=1/2$. The muon wavefunction is normalized according to
\begin{equation}\label{eq:munorm}
\int r^2(F^2+G^2)dr=1.
\end{equation}
We express the electron wavefunction as an expansion in partial waves,
according to
\begin{equation}
\varphi_e(\vec{r})=\sum_{\kappa\mu}a_{\kappa\mu t}\psi_{\kappa}^{\mu}=\sum_{\kappa\mu}a_{\kappa\mu t}\left(\begin{array}{c}g_{\kappa}\chi_{\kappa}^{\mu}\\if_{\kappa}\chi_{-\kappa}^{\mu}\end{array}\right),
\end{equation}
where $t$ is the $z$-component of the electron spin, and the $a_{\kappa\mu t}$
coefficients are given by
\begin{equation}\label{eq:apwe}
a_{\kappa\mu t}=i^{l_{\kappa}}\frac{4\pi}{\sqrt{2}}C\left(l_{\kappa}\frac{1}{2}j_{\kappa};\mu-t\, t\,
  \mu\right)Y_{l_{\kappa}}^{\mu-t\, *}(\hat{p}_e)e^{-i\delta_{\kappa}},
\end{equation}
with $\delta_{\kappa}$ the Coulomb phase shift (the distortion from a
plane wave due to the potential of the nucleus). The electron
wavefunctions are normalized in the energy scale, according to
\begin{equation}\label{eq:enorm}
\int d^3r\psi_{\kappa,W}^{\mu*}\psi_{\kappa',W'}^{\mu'}=2\pi\delta_{\mu\mu'}\delta_{\kappa\kappa'}\delta(W-W'),
\end{equation}
where $\psi_{\kappa,W}^{\mu}$ corresponds to a solution with energy
$W$.

\end{document}